\def\vev#1{\left< #1 \right>}     
\def\abs#1{\left| #1 \right|}     

\documentstyle[12pt]{article}

\begin{document}

\hsize=6in
\hoffset - .375 in

\begin{flushright}
TRI-PP-93-86\\
Nov. 1993
\end{flushright}

\renewcommand{\baselinestretch}{1.5} \small\normalsize

\

{\hskip -.375 in \centerline{\Large Moderate Supersymmetric $CP$ Violation}}

{\hskip -.375 in \centerline{Robert Garisto}}

{\hskip -.375 in \centerline{\it
TRIUMF,
4004 Wesbrook Mall,
Vancouver, B.C.,
V6T 2A3,
Canada
}}

\

\

It is well known that supersymmetry (SUSY) gives neutron and electron electric
dipole moments ($d_n$ and $d_e$) which are too large by about $10^{3}$.
If we assume a SUSY model cannot contain fine-tunings or large mass scales,
then one must require that the SUSY breaking mechanism give real
soft breaking parameters, in which case the minimal SUSY model has no
$CP$ violation other than from the CKM matrix (besides possible
strong $CP$ violating effects).
We show that in non-minimal SUSY models, a moderate amount of $CP$ violation
can be induced through one loop corrections to the scalar potential,
giving an effective phase of order $10^{-3}$, and thus implying $d_n$ and $d_e$
can be near their current experimental bounds $naturally$.
This moderate amount of SUSY $CP$ violation could also prove important
for models of electroweak baryogenesis.
We illustrate our results with a specific model.


\newpage
\section {Introduction}

Predictions for $CP$ violating effects in supersymmetric (SUSY) theories
have often been discussed with a certain ambiguity.
On the one hand, it is well known that when
the complex quantities in the theory are allowed to
have phases of order unity, the predicted neutron and electron electric dipole
moments ($d_n$ and $d_e$) are typically too large by perhaps $10^{3}$
\cite{hist dn,recent dn,Dugan etal,Kizukuri,arnowitt}.
In order to avoid this, the relevant quantities are often chosen
to be real, in which case the theory predicts no
non-Standard Model $CP$ violation (CPV) and negligible $d_n$ and $d_e$
\cite{foot strong}.
On the other hand, it has often been assumed that
the observation of $d_n$ around the current limit of $10^{-25}${e$\,$cm}
\cite{dnexpt} could easily be accommodated by a SUSY theory with
the phases somehow reduced by just the right amount.
These ideas are clearly in conflict: one cannot have a theory which
avoids fine-tunings by setting the SUSY parameters real,
and at the same time expect $d_n$ near its current upper bound.
The purpose of this paper is to describe a mechanism by which a moderate
amount of SUSY $CP$ violation can naturally appear in a theory
in which the soft SUSY breaking terms have been taken real.

The superpotential of the Minimal Supersymmetric Standard Model (MSSM) contains
the Yukawa sector of the theory, $W_Y$, and a Higgs mixing term,

\begin{equation}
W_{MSSM} = W_Y + \mu {\rm H}_u {\rm H}_d,
\label{MSSM superpot}
\end{equation}

\noindent
where ${\rm H}_u$  and ${\rm H}_d$ are Higgs doublet superfields.
If the soft breaking terms come from the superpotential, as in (\ref{L soft}),
then one can use ${\rm H}_u$  and ${\rm H}_d$ to rotate away the phase of
$\mu$.

In order to avoid an additional hierarchy problem brought on
by $\mu/M_{GUT} \ll 1$ \cite{Kim Nilles}, the MSSM is often extended
by adding a singlet superfield ${\rm N}$, whose
scalar component's vacuum expectation value (VEV) generates the Higgs
mixing term (see \cite{Ellis etal} and references therein).
We refer to this model as the N+MSSM.
One can use an $R$ symmetry to forbid $B$ and $L$ violating terms
in $W_Y$, and  to allow only cubic terms involving N,
so that the superpotential can be written as

\begin{equation}
W_{N+MSSM} = W_Y + h {\rm N} {\rm H}_u{\rm H}_d + a {\rm N}^3 .
\label{NMSSM superpot}
\end{equation}

\noindent
Note that we can use the Higgs and singlet ${\rm N}$
superfields to rotate away the
phases of $h$ and $a$.
This again assumes that
the soft SUSY breaking Lagrangian
can be written in the low energy supergravity (SUGRA)
parametrization \cite{Raby}:

\begin{equation}
-{\cal L}_{soft} = \abs{m_i}^2 \abs{\varphi_i}^2 +
\Bigl( {1 \over2} \sum_\lambda\tilde m_\lambda   \lambda \lambda  +
      \bar  A  \left[ W^{(3)} \right]_\varphi  +
      \bar  B  \left[ W^{(2)} \right]_\varphi  + h.c.\Bigr),
\label{L soft}
\end{equation}

\noindent
where $\varphi_i$ are the scalar superpartners, $\lambda$ are the
gauginos, and $[\ ]_\varphi$ means take the scalar part.
Here $W^{(2)}$ and $W^{(3)}$ are the quadratic and cubic pieces of the
superpotential, so that in the MSSM, $W^{(3)} = W_Y$, and
$W^{(2)}=\mu{\rm H}_u{\rm H}_d$; and in the N+MSSM, $W^{(3)}=W$.
We have defined the soft breaking parameters $\bar A\equiv A m_0^*$
and $\bar B\equiv B m_0^*$ to include a mass scale $m_0$.
The parameters $A$, $B$, their mass scale $m_0$,  and the
gaugino masses $\tilde m_\lambda$, can all be complex.
These parameters contribute to $d_n$ at the order of
$10^{-22} \tilde\varphi / \tilde M^2${e$\,$cm}, where $\tilde \varphi$ is a
combination of the phases of the parameters, and $\tilde M^2$ is
a combination of superpartner masses, normalized to the weak scale.
The only known ways to make such a large $d_n$ compatible with the experimental
upper bound
are to fine-tune the phase $\tilde \varphi$ to order  $10^{-3}$;
have superpartner masses of order a few TeV;
or somehow require all the phases to naturally be zero
\cite{SUSYdn}.
Both the first and second approach eliminate much of the
attractiveness of SUSY \cite{Kane etal}.
For example, having large superpartner
masses virtually eliminates the possibility of radiative breaking
of $SU_2\times U_1$, which was one of the major successes of SUSY.
Losing this is especially undesirable now that the top mass is large
enough to make it work.
So we will henceforth assume that
$A$, $B$, $m_0$ and $\tilde m_\lambda$ are all real
\cite{foot physical}.
We do $not$ have an explanation for how these conditions will be satisfied,
but merely state that any complete SUSY model which has superpartners
of order the weak scale must either satisfy these criteria, or provide an
explanation for how their phases could naturally be of order $10^{-3}$
\cite{foot high E spont}.

Imposing these `no fine-tuning criteria'
means that the only source of CPV in either
the MSSM or the N+MSSM is the CKM phase \cite{foot strong}.
CKM contributions to $d_n$ and $d_e$ from renormalization group running
\cite{Dugan etal} and from finite effects \cite{arnowitt,SUSYdn} are
below $10^{-30}$ {e$\,$cm}, and are thus unobservably small.
So if $d_n$ or $d_e$ were detected in the near future, could a SUSY theory
with superpartners of order the weak scale explain them
without resorting to fine-tunings?

There have also been some interesting models of baryogenesis at the electroweak
scale \cite{Turok Zadrozny,Dine BAU,Cohen Nelson},
which require CPV beyond the CKM phase \cite{foot Shaposhnikov}.
Could one construct a model with sufficient CPV for electroweak baryogenesis,
while satisfying the upper bounds on $d_n$ and $d_e$, without fine-tunings?

With these questions in mind, we describe a mechanism
by which a moderate amount of CPV can naturally arise in a
non-minimal SUSY theory
through loop corrections to the Higgs potential.
The idea is that
a phase which is unobservable at tree level can
introduce an observable effective phase
through loop effects.  This effective phase
will always be smaller than a tree level
observable phase because of the usual factors of suppression
associated with loops.
Such a phase can make moderate
contributions to $d_n$  and $d_e$, and may be useful in explaining the
observed baryon asymmetry.

In Section II, we present an illustrative model which provides an existence
proof for this moderate CPV mechanism.  In that section,
we show that at tree level in the scalar potential,
SUSY CPV is essentially unobservable.  In Section III, we show how this CPV
can appear in the observable sector in the one loop
effective potential.
The magnitude of this CPV will be suppressed by loop coefficients, so that
$d_n$ and $d_e$ can naturally be near their current
experimental bounds.

\section{The Model}

Let us construct a model which has complex couplings only to terms which
contain
singlet scalar fields which have zero VEVs. We also need these particles
to have no tree level couplings to quarks or leptons.
This means that the Higgs scalar potential will be $CP$ conserving, as will
all tree level vertices outside the neutral Higgs sector.
At this order, there is no one loop contribution to $d_n$ or $d_e$.  After one
loop corrections to the scalar potential ($V$),
a small phase can be induced into these vertices, which generates
moderate $d_n$ and $d_e$.
To do this in a model which is technically natural, one needs
to add at least two such singlets $(N',\ N'')$
to the N+MSSM.  In order that
they have zero VEVs, we impose a discrete symmetry on their superfields:
$({\rm N}',{\rm N}'') \rightarrow -({\rm N}',{\rm N}'')$.
Then the most general
cubic superpotential respecting this additional symmetry is:

\begin{equation}
W = W_Y +
h {\rm N} {\rm H}_u{\rm H}_d + a {\rm N}^3 + c' {\rm N}{{\rm N}'}^2 + c''
{\rm N}{{\rm N}''}^2
+ b {\rm N}{\rm N}'{\rm N}'' .
\label{model superpot}
\end{equation}

\noindent
One sees that the fields ${\rm N}'$ and ${\rm N}''$
have no direct couplings to quarks,
leptons, or gauge particles.  We will see below that they can each
have zero VEVs, and thus their couplings will not affect the tree level
minimum of the scalar potential.
They do not affect
$CP$ violating observables studied to date (at tree level in $V$),
so we term this sector {\it invisible}.
This is merely nomenclature.
It should be possible to detect these particles, and perhaps even to see
$CP$ violating effects directly in processes in which they are produced,
but they are certainly invisible when considering one loop processes
involving only external quarks, leptons, and gauge bosons.

Notice that we do not have enough freedom to rotate away all the phases
in (\ref{model superpot}), and that
after making the visible sector $CP$ conserving,
the reparametrization invariant $b^2 {c'}^* {c''}^*$ can be complex
\cite{foot alt choice}.
This is the phase which will produce CPV in the one loop scalar potential.
Our first task is to be sure that this phase
does not produce any CPV in the visible
sector at tree level in $V$, else $d_n$ will again be too large.

All supersymmetric contributions to $d_n$ come from the mass matrices
of squarks and gauginos---if the mass
matrices can all be made real, the SUSY contribution to $d_n$ disappears.
If they are complex, the gaugino-squark-quark couplings become
complex and contribute to $d_n$ through loop diagrams \cite{hist dn}.
Let us write the down squark mass matrix
in a partially diagonalized basis:
\begin{equation}
\pmatrix{ {\mu_{dL}}^2 {\bf 1} + \hat M_D^2 &
(\bar A^* - hn e^{i \theta_1}\, \tan\beta) \hat M_D\cr
(\bar A^* - hn e^{i \theta_1}\, \tan\beta)^* \hat M_D&
{\mu_{dR}}^2 {\bf 1} + \hat M_D^2\cr } , \label{MDsqk}
\end{equation}

\noindent
where $\hat M_D$ is the diagonal, real,
$N_F \times N_F$ quark mass matrix (where $N_F$ is the number of families),
and ${\mu_{q\, L,R}}^2 \sim \abs{m_{3/2}}^2$.
Here $n=\abs{\vev{N}}$ (so that $hn$ takes the place of $\mu$ of the MSSM),
and $\tan\beta$ is the ratio of Higgs VEVs.
The angle $\theta_1$ is one of the relative phases between the three
VEVs, and is defined in (\ref{theta defs}).
If (as we have assumed)
the soft breaking parameters are real, and the minimum of
the scalar potential $V$ is $CP$ conserving, then this matrix is real.


Next we can write the chargino mass matrix, $M_{\chi^+}$,
\begin{equation}
\pmatrix { \tilde m_{W} & g_2v_2 \cr
            g_2v_1      & h n e^{i \theta_1} \cr} ,
\label{Mchargino}
\end{equation}

\noindent
in the basis of \cite{HaK} (with the argument of the Higgs VEVs rotated
into $\theta_1$).
Here $\tilde m_{W}$ is
the $SU_2$ soft breaking gaugino mass, and
$g_2$ is the $SU_2$ coupling constant.
Again, if the minimum of $V$ is $CP$ conserving, then this matrix is real.

Since we have added three neutral fields to the MSSM (or two to the
N+MSSM), the neutralino mass matrix, $M_{\chi^0}$,
is $7\times7$.  We extend the basis of \cite{HaK} with $\psi_N,\ \psi_{N'},$
and $\psi_{N''}$:

\begin{equation}
\pmatrix{
 \tilde m_{B} &                   0 & -g_1v_1/\sqrt2 & g_1v_1/\sqrt2
&  0        & 0 & 0 \cr
                   0 &   \tilde m_{W} &   g_2v_1/\sqrt2    &-g_2v_2/\sqrt2
&  0        & 0 & 0 \cr
      -g_1v_1/\sqrt2 &  g_2v_1/\sqrt2 &                  0 &-h ne^{i\theta_1}
&  -h v_1 & 0 & 0 \cr
     g_1v_2/\sqrt2   & -g_2v_2/\sqrt2 & -h ne^{i\theta_1}&      0
&  -h v_2   & 0 & 0 \cr
  0            & 0                  & -h v_1       & -h v_2
& 3ane^{i(\theta_3- 2\theta_1)} & X  & Y   \cr
0&0&0&0& X  & c'n &  bn  \cr
0&0&0&0& Y  & bn  & c''n \cr
                                           },
\label{Mneutralino}
\end{equation}

\noindent
where $g_1$ is the $U_1$ coupling constant, and
$\tilde m_{B}$ is the $U_1$ gaugino mass.
The angle $\theta_3$ is also defined in (\ref{theta defs}).
The cross terms $X$ and $Y$, which mix $\psi_{N'}$ and $\psi_{N''}$
with the visible sector, are proportional to the VEVs of $N'$ and
$N''$, so that if these VEVs are zero,
$\psi_{N'}$ and $\psi_{N''}$
decouple from the visible sector.
The resulting $5\times5$ visible sector matrix is real, if
$\sin\theta_1=\sin\theta_3=0$.  In that case, all of the mass matrices are
real, and there is no new SUSY contribution to $d_n$.


To see if this is the case, we must consider the scalar potential $V$.
We define two Higgs doublets of the same hypercharge,
and their VEVs,
\begin{equation}
\vev{\phi_1} \equiv \vev{{H_d}^c} \equiv
{\left\lgroup \matrix{0 \cr v_1 \cr }\right\rgroup} ,\
\vev{\phi_2} \equiv \vev{H_u} \equiv
{\left\lgroup \matrix{0 \cr v_2 e^{i \xi}\cr}  \right\rgroup},
\label{defofphis}
\end{equation}

\noindent
and the VEVs of the singlet fields

\begin{equation}
\vev{N} \equiv n e^{i \varphi},\
\vev{N'} \equiv n' e^{i \varphi'},\
\vev{N''} \equiv n'' e^{i \varphi''}.
\end{equation}

\noindent
It turns out in this model that if $n'=n''=0$, there are
only three combinations of these VEV phases,

\begin{equation}
\theta_1 \equiv \xi + \varphi,\ \,
\theta_2 \equiv \theta_1 - \theta_3 =\xi - 2\varphi,\ \,
\theta_3 \equiv 3 \varphi,
\label{theta defs}
\end{equation}

\noindent
which appear in the tree level scalar potential.
Elsewhere, a more general linear combination of $\theta_1$ and $\theta_3$
(with integer coefficients) can appear.

Let us write the scalar potential for our model,

\begin{eqnarray}
V &&\!\!\!\!\!= (h^2 \abs{N}^2 + m_1^2) \abs{\phi_1}^2 +
(h^2 \abs{N}^2 + m_2^2) \abs{\phi_2}^2
\nonumber\\
&&\!\!\!\!\!- \left[ \left( \bar A h N + 3 a h {N^*}^2 + h {c'}^* {N'}^{*2}
+ h {c''}^* {N''}^{*2} \right) (\phi_1^\dagger \phi_2) + H.c. \right]
\nonumber\\
&&\!\!\!\!\!+\lambda_1 (\phi_1^\dagger \phi_1)^2
+ \lambda_2 (\phi_2^\dagger \phi_2)^2 +
\lambda_3 (\phi_1^\dagger \phi_1)(\phi_2^\dagger \phi_2) +
\lambda_4 (\phi_1^\dagger \phi_2)(\phi_2^\dagger \phi_1)
\nonumber\\
%
&&\!\!\!\!\!+ m_0^2 \abs{N}^2 + m_0^2 \abs{N'}^2 + m_0^2 \abs{N''}^2
+ 9a^2 \abs{N}^4 + \abs{c'}^2 \abs{N'}^4 + \abs{c''}^2 \abs{N''}^4 +
\label{scalar potential}\\
&&\!\!\!\!\!+ (4 \abs{c'}^2 + \abs{b}^2) \abs{NN'}^2
+ (4 \abs{c''}^2 + \abs{b}^2) \abs{NN''}^2
\nonumber\\
&&\!\!\!\!\!+ \Bigl[ \bar A a N^3 + \bar A c' N {N'}^2 + \bar A c'' N {N''}^2
+ \bar A b NN'N''
\nonumber\\
&&\!\!\!\!\!+ 2(b^* c' + b {c''}^*) \abs{N}^2 N' {N''}^*
+ 3 ac' {N^*}^2 {N'}^2 + 3 ac'' {N^*}^2 {N''}^2
+ c' {c''}^* {N'}^2 {N''}^{*2}  + H.c. \Bigr]
\nonumber
\end{eqnarray}

\noindent
where \cite{GaH}

\begin{equation}
\lambda_1=\lambda_2 = (g_2^2 + g_1^2)/8,\ \lambda_3 = (g_2^2 - g_1^2)/4,\
\lambda_4= h^2 - g_2^2/2,
\end{equation}

\noindent
which is just the scalar potential for the N+MSSM
\cite{Ellis etal} plus terms which involve $N'$ and $N''$.
The minimum of $V$ can be written as

\begin{eqnarray}
\vev{V} &= \vev{V_{N+MSSM}} + \, K_{20} {n'}^2 + K_{11} n'n'' + K_{02} {n''}^2
&\nonumber\\
&+K_{40} {n'}^4 + K_{22} {n'}^2{n''}^2 + K_{04} {n''}^4,
&\label{vev V}
\end{eqnarray}

\noindent
where $K_{ij}$ depend upon all the other parameters.
One can show that for any choice
of the parameters in $V_{N+MSSM}$,
there exists a set of $\{b,c',c''\}$ such
that $n'=n''=0$ is a true minimum.  We will assume that this condition
is satisfied, so that $\vev{V} = \vev{V_{N+MSSM}}$.

Finally we must be sure there is no problem with spontaneous CPV.  As we said,
the potential depends only upon the three angles $\theta_{{\rm 1-3}}$
(only two of which are independent).  We can write

\begin{equation}
\vev{V} = \alpha_0 - \alpha_1 \cos\theta_1
- \alpha_2 \cos\theta_2
- \alpha_3 \cos\theta_3,
\end{equation}

\noindent
where the $\alpha_i$ are functions of the magnitudes of the three VEVs.
Differentiating with respect to $\theta_1$ and $\theta_3$, we see that
one solution to $\vev{V'}=0$ is $\sin\theta_i=0$.
Rom$\tilde{\rm a}$o \cite{Romao} showed that for this potential
($i.e.$ $\vev{V_{N+MSSM}}$), this is the only stable minimum---that
the spontaneous $CP$ violating solution is actually a saddle point.
Babu and Barr \cite{Babu Barr} make the interesting
claim that this can be made into a minimum by
large radiative corrections to the Higgs mass matrix, but these
require very heavy squark masses (in which case hard CPV need not be
suppressed by fine-tuning \cite{Kizukuri}), small charged Higgs mass, and
$\tan\beta \sim {\cal O}(1)$.  These conditions make satisfying the
CLEO bound on $b \rightarrow s \gamma$ nearly impossible \cite{my BSG}.
It is also unlikely that a model satisfying these conditions
could be consistent with such things as
Grand Unification and solutions to the Dark Matter problem \cite{Kane etal}.
Anyway, we can certainly choose parameters such that the minimum
of $V$ is $CP$ conserving, and such that $n'=n''=0$, so that all
the SUSY mass matrices (\ref{MDsqk})-(\ref{Mneutralino})
are real at tree level.

\section{A Loop Induced Observable Phase}

It would seem that since the tree level potential is $CP$ conserving,
one could not have a one loop potential which is $CP$ violating.
The important point to remember is that even though the {\it visible} sector
has
no CPV, there are still $CP$ violating couplings to $N'$ and $N''$.
Consider, for example, Figure 1, which gives a purely finite
contribution to a new term in $V$,
$\delta \lambda_5 (\phi_1^\dagger \phi_2)^2 + H.c$
(this term is not present in the tree level potential, so it must be finite).
The vertices are proportional to ${c'}^*$ and ${c''}^*$, and
the mixing between $N'$ and $N''$ contains pieces proportional to $b$.
Thus $\delta \lambda_5 \sim b^2 {c'}^* {c''}^*$, which has a
reparametrization invariant phase.  For $b$, $c'$, $c''$ of
order $1/2$, $\delta \lambda_5$ can be of order $10^{-3}$.

Actually, the operator (to which Figure 1 contributes) is more
accurately written as $k \vev{N}^2 (\phi_1^\dagger \phi_2)^2$,
which gives a contribution to $\vev{\delta V}$ of
$2 \abs{k} n^2 v_1^2 v_2^2 \cos(2 \theta_1 + {\rm Arg} k)$, where
${\rm Arg} k$ is
just $\theta_{CP} \equiv {\rm Arg}(b^2 {c'}^* {c''}^*)$.
We can write the general correction to $\vev{V}$ as

\begin{equation}
\vev{\delta V} = \sum_{x,y,z}^{integers}
\kappa_{x,y,z} \cos(x \theta_1 + y \theta_3 + z \theta_{CP}),
\label{delta V}
\end{equation}

\noindent
where the $\kappa_{x,y,z}$ are real coefficients, with the subscripts
$x,\ y,\ z$ taking on all integral values, though the $\kappa_{x,y,z}$
become negligible for large integers.
Since our model is renormalizable \cite{renorm} and $\vev{V}$ is $CP$
conserving at tree level, all one loop terms in (\ref{delta V}) with
$z \neq 0$ must be finite.
Note that Figure 1 gives a finite contribution to (\ref{delta V})
with $(x,y,z) = (2,0,1)$.
%

The perturbation in (\ref{delta V}) means that $\sin\theta_i = 0$ is no longer
a solution to $\vev{V(\theta_1,\theta_3)'}=0$.
Since the $\kappa_{x,y,z}$ are small,
the solution will lie close to this, so we can define
$\theta_i = {\theta_i}_0 + \varepsilon_i$, where $\sin{\theta_i}_0=0$.
The minimization condition can then be written in terms of the hessian,
and the perturbation:

\begin{equation}
\pmatrix
{ \partial^2 V/\partial \theta_1^2 &
\partial^2 V/\partial \theta_1 \partial\theta_3 \cr
  \partial^2 V/\partial \theta_1 \partial\theta_3 &
\partial^2 V/\partial\theta_3^2 \cr}
\pmatrix { \varepsilon_1 \cr \varepsilon_3 \cr}
          \simeq
\pmatrix {
\sum_{x,y,z}^{integers} x
\kappa_{x,y,z} \sin(x {\theta_1}_0 + y {\theta_3}_0 + z \theta_{CP}) \cr
\sum_{x,y,z}^{integers} y
\kappa_{x,y,z} \sin(x {\theta_1}_0 + y {\theta_3}_0 + z \theta_{CP}) \cr
}
\label{epsilon eqs}
\end{equation}

\noindent
and solved for $\varepsilon_i$.
Note that $\sin(x {\theta_1}_0 + y {\theta_3}_0 + z \theta_{CP}) =$
$\pm \sin(z \theta_{CP})$.
To find the effective $CP$ violating coefficient,
recall that $d_n$ gets a contribution from the imaginary part of
left--right squark mixing, which goes as $\sin\theta_1 \simeq \varepsilon_1$.
There will be finite contributions to $\varepsilon_1$ from several terms
in $\delta V$, but they will be of the same order or smaller than
that of $\delta \lambda_5$ from Figure 1.
{}From (\ref{epsilon eqs}), one finds that Figure 1 gives
$\varepsilon_1 \sim \abs{\delta\lambda_5} {v^2 \over \bar A \mu}
 \sin{2\beta} \sin\theta_{CP}$.
The squark mixing, gluino mediated contribution to $d_n$ \cite{SUSYdn} due to
this $\varepsilon_1$ can be written as

\begin{equation}
d_n \simeq 10^{-22} {\rm e\, cm}\ {\left( {100 \mbox{GeV} \over \tilde M}
\right)}^2
\abs{\delta \lambda_5}
{v^2 \over A m_0^2}
\sin\theta_{CP}   ,
\label{dn expr}
\end{equation}

\noindent
where we have defined a SUSY mass scale

\begin{equation}
\tilde M^2 \equiv {\tilde m_d^4 \over m_0 \, \tilde m_g} \ .
\label{overallMsusy}
\end{equation}

\noindent
There is a similar neutralino mediated contribution to $d_e$ which is
suppressed by $m_e/m_d$, and $\alpha_w/\alpha_s$,
but enhanced by the fact that sleptons tend to be lighter than squarks.
If we take
$A=1$, colored superpartners $\sim 300$GeV, sleptons $\sim 150$GeV, and
all other superpartners $\sim 100$GeV, one can have

\begin{eqnarray}
&&d_n \sim 10^{-26} \, \sin\theta_{CP} \, {\rm e\, cm} ,
\label{dn est}\\
&&d_e \sim 10^{-27} \, \sin\theta_{CP} \,
{\rm e\, cm} .
\label{de est}
\end{eqnarray}

\noindent
These estimates depend upon the parameters and the mass scales in the
theory, but the point is that the contributions entering at one loop
are naturally much smaller than
those from SUSY phases which contribute through tree level vertices.

\section{Concluding Remarks}

We have considered supersymmetric models which avoid excessively large
contributions to $d_n$ and $d_e$ by requiring the `no fine-tuning criteria' to
be satisfied, $i.e.$ that $A$, $B$, $m_0$ and $\tilde m_\lambda$ must be
real \cite{SUSYdn}.
We showed that it is possible for moderate
$CP$ violating effects to be induced at one loop in models which
have singlets with zero VEVs.  We used an illustrative model with
superpotential (\ref{model superpot}) and found that a stable minimum
exists at $\vev{N'}=\vev{N''}=0$.  This means that using the
tree level scalar potential, no $CP$ violating effects would
be detectable in conventional $CP$ violating observables because all
of the SUSY mass matrices are real.
We demonstrated that this model introduces small $CP$ violating phases
into the one loop effective potential, so that one is left with a moderate
contribution to $d_n$ and $d_e$.  One could easily have $d_n$ and $d_e$
near their
current experimental bounds in such a model, without the need for fine-tuning
or large superpartner mass scales.

Note that SUSY contributions to $d_n$ from
three gluon operators \cite{three gluon} do not affect our conclusions.
Assuming that the no fine tuning criteria are satisfied, such operators
will also give a negligible contribution to $d_n$.  After one loop corrections
to $V$ in our model, there will be  small contributions to $d_n$ from these
operators, but they will probably be smaller in magnitude than the
quark EDM contribution \cite{arnowitt}.
Thus (\ref{dn est}) and (\ref{de est}) are reasonable estimates of the
natural size of SUSY CPV possible in a model such as ours.

We have discussed the issue of {\it spontaneous} CPV in Section II
in the context of our model and concluded that we can easily choose
the minimum of $V$ to be $CP$ conserving.  It is worth noting that
Maekawa \cite{Maekawa} considered generating spontaneous CPV
at one loop in the MSSM, though Pomarol \cite{Pomarol one} showed
that such a model requires a $CP$ odd Higgs which is too light.
Pomarol also made the interesting point that a N+MSSM model
(which does not rule out ${\rm H}_u{\rm H}_d$, ${\rm N}$, or ${\rm N}^2$
terms by a symmetry)
with a strictly
$CP$ conserving Lagrangian might violate $CP$ spontaneously at tree level
with a phase of order $10^{-2}$, and might be able to explain the
$\varepsilon$ parameter as well as give $d_n$ near the current experimental
bound \cite{Pomarol two}.
The trouble is that the fine-tuning needed by such a model of spontaneous
CPV is actually much worse than that for hard CPV because the condition
which must be satisfied is of the form

\begin{equation}
\cos\theta = \abs{{X \over Y}} \simeq 1 - {1\over 2} \theta^2,
\end{equation}

\noindent
where $\theta$ (or $\pi - \theta$) is the relevant spontaneous CPV phase,
and $X$ and $Y$ are some combination of parameters and VEVs.
We need $\theta$ to be small to satisfy the bound on $d_n$,
which we can achieve only if $\delta \equiv (Y-X)/Y$ is of order $\theta^2$.
For example, if we need $\theta\sim 10^{-2}$, then $\delta$ must
be fine-tuned to be of order $10^{-4}$, which is completely unacceptable.

As we alluded to in the Introduction, having
a moderate amount of CPV is necessary in models which generate the
baryon asymmetry at the electroweak scale \cite{foot Shaposhnikov}.
A recent interesting model of electroweak baryogenesis used
CPV from the Higgs scalar mixing coefficient $\mu_{12}^2$
\cite{Turok Zadrozny},
which can be defined as the coefficient of the $\vev{\phi_1^\dagger \phi_2}$
term in $\vev{V}$.
It was pointed out \cite{Dine BAU} that  $\mu_{12}^2$
can be rotated out of the Higgs
potential, but the resulting phase which appears in the gaugino mass matrices
was then used by \cite{Cohen Nelson}.  They found that with the small phase
allowed by the limit imposed by $d_n$, there is probably
sufficient CPV for the observed baryon asymmetry.
Our results change these conclusions in two ways.  At tree level, there
is {\it no} phase in the gaugino mass matrices (after imposing
the no fine-tuning criteria),
and no way for ${\rm Arg}\mu_{12}^2$ to cause CPV.
Then, using the one loop effective potential in a model such as ours,
there can be an effective phase $\varepsilon_1$
($\equiv \theta_1 - {\theta_1}_0$)
introduced into the
gaugino mass matrix of order
$\varepsilon_1 \sim 10^{-3} \theta_{CP}$.
Although this is a large suppression, $\theta_{CP}$ can be of order
unity and so $\varepsilon_1$ should generate the same level of CPV
as the phase used in  \cite{Cohen Nelson}, which was bounded by $d_n$ anyway
\cite{foot CaN}.

{}From the standpoint of explaining the baryon asymmetry at the
electroweak scale, or $d_n$ and $d_e$ near their current experimental
bounds, a loop induced observable phase
provides an attractive alternative to
the fine-tuning needed in the MSSM.
If $d_n$ or $d_e$ were observed in the near future,
and if superpartners were determined to be of order the weak scale,
SUSY model builders would have to appeal to a mechanism such
as ours, which naturally explains small effective SUSY phases.

\

{\hskip -.375 in \centerline{ACKNOWLEDGEMENTS}}

I sincerely appreciate helpful comments from
Gordy Kane, who was also involved in the early stages of this work.
I also appreciate helpful conversations with H. Haber and J. Soares.
This work was supported in part by a grant from the National Sciences
and Engineering Research Council of Canada.





\newpage

{\hskip -.375 in \centerline{FIGURE CAPTION}}

Fig 1: One loop contribution to the operator
$\delta \lambda_5 (\phi_1^\dagger \phi_2)^2$.
The `X' indicates $N'$--$N''$ mixing, which is required if
$\delta \lambda_5$ is to contain a reparametrization invariant phase.


\begin{thebibliography}{99}
\bibitem{hist dn}J. Polchinski \& M. Wise, Phys. Lett. {\bf B125}, 393 (1983);
F. del Aguila, M. Gavela, J. Grifols \& A. Mendez, Phys. Lett. {\bf B126}, 71
(1983);W. Buchmuller \& D. Wyler, Phys. Lett. {\bf B121}, 321 (1983).
%
\bibitem{recent dn}
W. Fischler, S. Paban, S. Thomas, Phys. Lett. {\bf B289}, 373 (1992);
P. Nath, Phys. Rev. Lett. {\bf 66}, 2565 (1991);
L. Hall \& L. Randall, Nucl. Phys. {\bf B352}, 289 (1991);
R. Mohapatra in {\it CP Violation}, edited by C. Jarlskog,
World Scientific, Singapore (1989);
I. Bigi \& F. Gabbiani, Notre Dame preprint UND-HEP90-BIG06;
T. Kurimoto, Prog. Theor. Phys., {\bf 73}, 209 (1985);
M. Vysotsky, Sov. Phys. Usp, {\bf 28}, 667 (1985);
J. G\' erard, W. Grimus, A. Raychaudhuri, \& G. Zoupanos,
Phys. Lett. {\bf B140}, 349 (1984).
%
\bibitem{Dugan etal}
M. Dugan, B. Grinstein \& L. Hall, Nucl. Phys. {\bf B255}, 413 (1985).
%
\bibitem{Kizukuri}Y. Kizukuri \& N. Oshimo, Phys. Rev. {\bf D46}, 3025 (1992).

\bibitem{arnowitt}
R. Arnowitt, M. Duff, \& K. Stelle, Phys. Rev. {\bf D43}, 3085 (1991);
R. Arnowitt, J. Lopez, \& D. V. Nanopoulos, Phys. Rev.
{\bf D42}, 2423 (1990).
%

\bibitem{foot strong}We will ignore possible strong CPV effects.
While $d_n$ depends on the amount of strong CPV,
a non-negligible $d_e$ could not be explained by CKM and strong
CPV effects alone.


\bibitem{dnexpt}K. Smith $et\ al$, Phys. Lett. {\bf B234}, 191 (1990).

\bibitem{Kim Nilles}J. Kim \& H. Nilles, Phys. Lett. {\bf B138}, 150 (1984).


%
\bibitem{Ellis etal}
J. Ellis, J. Gunion, H. Haber, L. Roszkowski, F. Zwirner,
Phys. Rev. {\bf D39}, 844 (1989).

\bibitem{Raby}S. Raby, Santa Cruz '86 TASI proceedings (1986).


\bibitem{SUSYdn} For a complete statement of the criteria,
see R. Garisto, TRI-PP-93-24 (1993) hep-ph/9306318.


\bibitem{Kane etal}G. Kane, C. Kolda, L. Roszkowski \& J. Wells,
University of Michigan preprint UM-TH-93-24 (1993).


\bibitem{foot physical}There are only two new physical phases in these
parameters, but if the physical phases are zero, we can always write the
theory such that $A$, $B$, $m_0$ and $\tilde m_\lambda$ are real
\cite{SUSYdn}.




\bibitem{foot high E spont}It may be possible to introduce small phases at the
weak scale by a spontaneous symmetry breaking at high energy.  See
M. Dine, R. Leigh, \& A. Kagan,
Phys. Rev. {\bf D48}, 4269 (1993);
and
K. Babu \& S. Barr, Bartol Research Inst. preprint BA-93-48 (1993)
hep-ph/9309249.





\bibitem{Turok Zadrozny}N. Turok \& J. Zadrozny, Nucl. Phys. {\bf B369}, 729
(1992).
%
\bibitem{Dine BAU}M. Dine, P. Huet \& R. Singleton Jr., Nucl. Phys.
{\bf B375}, 625 (1992).
%
\bibitem{Cohen Nelson}A. Cohen \& A. Nelson,
Phys. Lett. {\bf B297}, 111 (1992);
see also A. Cohen, D. Kaplan, A. Nelson,
Phys. Lett. {\bf B294}, 57 (1992).



\bibitem{foot Shaposhnikov}
There is a recent claim by G. Farrar and M. Shaposhnikov,
Phys. Rev. Lett. {\bf 70}, 2833 (1993), that the CKM phase
is sufficient for electroweak baryogenesis, but the general consensus is that
CPV from the CKM phase is insufficient to explain the observed asymmetry.


\bibitem{foot alt choice}
For simplicity we choose the basis with $h$ and $a$ real.
If one chooses to keep the phase in $a$ instead, one can show that
observable CPV still occurs only at one loop in $V$,
and the results are the same as in our convention.


%
%

\bibitem{HaK}H. Haber \& G. Kane, Phys. Repts. {\bf 117}, 75 (1985).
\bibitem{GaH} J. Gunion \& H. Haber, Nucl. Phys. {\bf B272}, 1 (1986).
%



\bibitem{Romao}J. Rom$\tilde{\rm a}$o, Phys. Lett. {\bf B173}, 309 (1986).
\bibitem{Babu Barr}K. Babu \& S. Barr, Bartol Research Inst. preprint
BA-93-42 (1993) hep-ph/9308217.


\bibitem{my BSG}R. Garisto \& J. N. Ng, Phys. Lett. {\bf B315}, 372 (1993).


\bibitem{renorm}Our model has a complete, $N=1$ supersymmetric Lagrangian,
and is thus renormalizable (see for example M. Sohnius, Phys. Reports
{\bf 128}, 2 (1985)).

%
\bibitem{three gluon}
S. Weinberg, Phys. Rev. Lett. {\bf 63}, 2333 (1989);
J. Dai, H. Dykstra, R. Leigh, S. Paban \& D. Dicus, Phys. Lett. {\bf B237}, 216
(1990).

\bibitem{Maekawa} N. Maekawa, Phys. Lett. {\bf B282}, 387 (1992).
\bibitem{Pomarol one} A. Pomarol, Phys. Lett. {\bf B287}, 331 (1992).
\bibitem{Pomarol two} A. Pomarol, Phys. Rev. {\bf D47}, 273 (1992).

%
\bibitem{foot CaN}The phase of $\mu$ used by \cite{Cohen Nelson} does not
contribute to CPV unless $B m_0^*$ is complex and fine-tuned to
satisfy the bound on $d_n$.
\end{thebibliography}
\end{document}